\documentclass[aps,prb,twocolumn, reprint]{revtex4-2}
\usepackage[pdftex, colorlinks=true, linkcolor=blue, citecolor=blue, urlcolor=blue]{hyperref}
\usepackage{graphicx}
\usepackage{aurical}
\usepackage{amsmath, amssymb}
\usepackage[T1]{fontenc}
\usepackage[version=3]{mhchem}
\usepackage{braket}
\graphicspath{{./figures/}}
\DeclareGraphicsExtensions{.png,.pdf}

\newcommand{\una}{Universit\'{e} de Nouakchott, Facult\'{e} des Sciences et Techniques, D\'{e}partement de Physique, Avenue du Roi Fai\c{c}al, 2373, Nouakchott, Mauritania}

\begin{document}

\title{An Energy Integration Free Kubo-Bastin Formula Decomposition}

\author{O. Ly}
\email{ousmanebouneoumar@gmail.com}
\affiliation{\una}

\begin{abstract}
Kubo formulae play a central role in modern spintronics and condensed matter physics, serving as the foundational ground for studying transport responses in the linear regime. In this work, we propose a reformulation of the widely used Kubo-Bastin decompositions that eliminates the need for numerical energy integration. By performing these integrations analytically for generic periodic systems, our approach significantly reduces computational cost and simplifies the calculation of transport coefficients, while maintaining a clear distinction between the relevant Fermi-surface and Fermi-sea transport components. { Our formulation suggests that the Fermi-surface term can be attributed to a pure intraband contribution.} {In contrast to widely used energy integration free expressions that are strictly limited to zero temperature, the present formulation remains valid at arbitrary temperatures.}
\end{abstract}

\maketitle

\section{Introduction}
The derivation of linear response coefficients typically begins with the foundational Kubo formula \cite{Kubo1956}. A widely adopted refinement was later introduced by Bastin et al. \cite{Bastin1971}, which expresses response coefficients in terms of the Green functions of the system's Hamiltonian. This Kubo-Bastin framework is frequently partitioned into "surface" and "sea" contributions, a technique originally proposed by Smr\v{c}ka and St\v{r}eda \cite{Smrcka1977}. More recently, this decomposition was revised by Bonbien and Manchon \cite{Bonbien2020}, who provided a corrective formulation that explicitly separates an overlap term previously embedded in the initial decomposition.
{This specific decomposition has since been widely adopted across various contexts, including orbital pumping \cite{Go2025, Han2025}, valley Hall responses \cite{Aktor2021}, and transport phenomena mediated by Rashba coupling \cite{Xie2022, Pezo2025, Pezo2026}. Furthermore, it has been successfully applied to graphene \cite{Sousa2024, Medina2026} and related Van der Waals heterostructures \cite{Sousa2024}.}

A common challenge across these formulations is the requirement for an integration over the energy spectrum in addition to the summation over momentum space. For large-scale systems, this dual integration can become computationally prohibitive, often necessitating the use of methods like the Kernel Polynomial Method (KPM) to remain tractable \cite{Garcia2015, Canonico2024}. 

In parallel to the Green function formalism, it is possible to derive the Kubo formula entirely within momentum space, bypassing the need for numerical energy integration. This was demonstrated in the seminal work by Cr\'{e}pieux and Bruno \cite{Crepieux2001}, which established a formal correspondence between the Green function approach and an energy-integration-free representation. However, the Green function formalism remains popular because it enables a direct separation between surface and sea contributions, providing physical insight into the origins of various transport phenomena.

In this work, we demonstrate that this physically intuitive decomposition does not require numerical energy integration. By expanding the decomposition in the system's eigenbasis, we show that the energy integrals can be performed analytically {at finite temperature}. This results in simplified expressions for both surface and sea contributions, requiring only a single integration over momentum space at the underlying chemical potential. {Eliminating this energy integration certainly drastically reduces the computational cost associated with the underlying response formulae.}

{It is worth emphasizing that a substantial simplification of the Kubo-Bastin formula terms has been previously achieved by Freimuth \textit{et al}~\cite{Freimuth2014} in the  constant broadening approximation, however assuming zero temperature. Within such a framework, the response tensors can be written without explicit energy integrals, categorized either by their symmetry properties~\cite{Zelezny2014, Zelezny2017, Freimuth2017, Gonzalez2020, Lee2022, Cui2023, Johansson2024, Hu2025, Lai2025, Zhang2025, Hirakida2026, Sim2026} or via a decomposition into distinct inter and intraband contributions~\cite{Li2015, Go2020a, Go2020b, Han2025b, Tenzin2025}, which can be also seen as a decomposition into surface and sea terms \cite{Lee2026}. However, this simplification relied entirely on the reduced form of the Fermi-Dirac function and its energy derivative at \textit{zero temperature}. By contrast, our energy integration-free decomposition holds for \textit{arbitrary temperatures}.}

\section{Surface and Sea Terms of the Kubo-Bastin Formula}
\label{sec:formulas}
Le us consider a generic system, with Hamiltonian $H$. One can define the retarded (advanced) Green functions at transport energy $\varepsilon$ as 
\begin{eqnarray}
    G^{r(a)}=\frac{1}{\varepsilon-H \pm i\eta},
\end{eqnarray}
where $\eta$ is an infinitesimal real number.
The Kubo-Bastin formula for the linear response correlation of two general operators, $\mathcal{O}^{\alpha}$ and $\mathcal{O}^{\beta}$, is given in terms of these Green functions according to \cite{Bastin1971}:
\begin{equation}
    \label{eq:kb}
    \sigma_{\alpha\beta} = -2\hbar\int d\varepsilon f(\varepsilon) \text{Tr} \left[ \rm{Im}(\mathcal{O}^{\alpha}\partial_{\varepsilon} G^r\mathcal{O}^{\beta})\delta(\varepsilon-H) \right].
\end{equation}
{{Where $f(\varepsilon)$ is the Fermi-Dirac distribution function.}}
For a Bloch Hamiltonian, a summation over momentum space is assumed.

The Smr\v{c}ka-St\v{r}eda decomposition \cite{Smrcka1977} partitions this response into two terms, $\sigma_{\alpha\beta}^{\mathrm{I}}$ and $\sigma_{\alpha\beta}^{\mathrm{II}}$, which can be written as:
\begin{equation}
    \label{eq:sI}
    \sigma_{\alpha\beta}^{\mathrm{I}} = -i{\hbar}\int d\varepsilon f'(\varepsilon) \text{Tr} \left[ \mathcal{O}^{\alpha} G^r\mathcal{O}^{\beta} \delta(\varepsilon-H) \right],
\end{equation}
and 
\begin{equation}
    \label{eq:sII}
    \sigma_{\alpha\beta}^{\mathrm{II}} = \frac{\hbar}{2\pi}\int d\varepsilon f(\varepsilon) \text{Tr} \left[ \mathcal{O}^{\alpha} G^r\mathcal{O}^{\beta} \partial_{\varepsilon} G^r - \mathcal{O}^{\alpha} \partial_{\varepsilon} G^r \mathcal{O}^{\beta} G^r \right].
\end{equation}

Alternatively, using the recently introduced permutation decomposition \cite{Bonbien2020}, the Kubo-Bastin formula can be expressed as the sum of two modified terms $\tilde{\sigma}_{\alpha\beta}^{\mathrm{I}}$ and $\tilde{\sigma}_{\alpha\beta}^{\mathrm{II}}$, which here are expressed in terms of $\delta(\varepsilon-H)$ as:
\begin{equation}
    \label{eq:BMI}
    \tilde{\sigma}_{\alpha\beta}^{\mathrm{I}} = -\pi{\hbar}\int d\varepsilon f'(\varepsilon) \text{Tr} \left[ \mathcal{O}^{\alpha}\delta(\varepsilon-H)\mathcal{O}^{\beta} \delta(\varepsilon-H) \right],
\end{equation}
and 
\begin{equation}
     \label{eq:BMII}
    \tilde{\sigma}_{\alpha\beta}^{\mathrm{II}} = -i{\hbar}\int d\varepsilon f(\varepsilon) \text{Tr} \left[ \mathcal{O}^{\alpha}\delta(\varepsilon-H)\mathcal{O}^{\beta}\partial_\varepsilon (G^r+G^a) \right].
\end{equation}
By introducing a surface-sea response overlap term,
\begin{equation}
     \label{eq:ol}
    \sigma_{\alpha\beta}^{ol} = \frac{\hbar}{2}\int d\varepsilon f'(\varepsilon) \text{Tr} \left[ \rm{Im}(\mathcal{O}^{\alpha}(G^r+G^a)\mathcal{O}^{\beta})\delta(\varepsilon-H) \right],
\end{equation}
one can relate these two decomposition schemes as follows:
\begin{equation}
\label{eq:fromOverlap}
\begin{aligned}
    \tilde{\sigma}_{\alpha\beta}^{\mathrm{I}} &= \sigma_{\alpha\beta}^{\mathrm{I}} - \sigma_{\alpha\beta}^{ol}, \\
    \tilde{\sigma}_{\alpha\beta}^{\mathrm{II}} &= \sigma_{\alpha\beta}^{\mathrm{II}} + \sigma_{\alpha\beta}^{ol}.
    \end{aligned}
\end{equation}
The terms $\tilde{\sigma}_{\alpha\beta}^{\mathrm{I}}$ and $\tilde{\sigma}_{\alpha\beta}^{\mathrm{II}}$ are referred to as the true surface and sea contributions \cite{Bonbien2020}. In cases where the overlap term vanishes, the permutation decomposition coincides with the original Smr\v{c}ka-St\v{r}eda formulation.

\section{Energy-Integration-Free Expressions for response coefficients}

To simplify the evaluation of relevant conductivity terms, we reformulate the expressions presented above, in the system's eigenbasis. 
This allows us to evaluate the energy integrals analytically. 
We generally write the responses in terms of a momentum dependent kernel matrix $\mathcal{K}(\boldsymbol{k})$, 
\begin{equation}
    \label{eq:generalResponse}
    \sigma_{\alpha\beta}=\sum_{mn, \boldsymbol{k}} \mathcal{O}^{\beta}_{mn}(\boldsymbol{k}) \mathcal{O}^{\alpha}_{nm}(\boldsymbol{k})  \mathcal{K}_{mn}(\boldsymbol{k}),
\end{equation}
where $\mathcal{O}^{\beta}_{mn}$ is the matrix element of the operator $\mathcal{O}^{\beta}$ between two eigenstates $\ket{m}$ and $\ket{n}$. {The summation over $\boldsymbol{k}$ is the usual momentum space summation over the first Brillouin zone. }

{ In what follows we derive the different kernels $\mathcal{K}_{mn}(\boldsymbol{k})$ to be considered for the relevant transport coefficients discussed in Sec. \ref{sec:formulas}.}
 
The full Kubo-Bastin kernel can be expressed as \cite{Crepieux2001} (see Appendix~\ref{appa:full}):
\begin{equation}
    \label{eq:fullKernel}
    \mathcal{K}_{mn}(\boldsymbol{k}) = -2i\hbar  \frac{f(\varepsilon_n)}{(\varepsilon_{nm} + i\eta)^2},
\end{equation}
with $\varepsilon_{nm}=\varepsilon_n - \varepsilon_m$, {while $\varepsilon_m$ is the $m$th eigen-energy of the Hamiltonian}.

Similarly, the individual terms $\sigma_{\alpha\beta}^{\mathrm{I}}$ (see Appendix \ref{appa:sI}) and $\tilde\sigma_{\alpha\beta}^{II}$ (see Appendix \ref{appa:bmII})  have the following kernels:
\begin{equation}
    \label{eq:stredaIKernel}
    \mathcal{K}^{\mathrm{I}}_{mn}(\boldsymbol{k}) = -i\hbar   \frac{f'(\varepsilon_n)}{\varepsilon_{nm} + i\eta}, 
\end{equation}
\begin{equation}
    \label{eq:bmIIKernel}
    \tilde{\mathcal{K}}^{\mathrm{II}}_{mn}(\boldsymbol{k}) = 2i\hbar f(\varepsilon_m) \frac{\varepsilon_{nm}^2 - \eta^2}{(\varepsilon_{nm}^2 + \eta^2)^2}.
\end{equation}
For the overlap term, the kernel reads (see Appendix \ref{appa:ol}):
\begin{equation}
    \label{eq:olKernel}
    \mathcal{K}^{ol}_{mn}(\boldsymbol{k}) = -i\hbar   \frac{\varepsilon_{nm}f'(\varepsilon_n)}{\varepsilon_{nm}^2 + \eta^2}.
\end{equation}
The remaining components, $\sigma_{\alpha\beta}^{\mathrm{II}}$ and $\tilde{\sigma}_{\alpha\beta}^{\mathrm{I}}$, can be straightforwardly deduced from Eq. (\eqref{eq:fromOverlap}). 
{Particularly, the term $\tilde{\sigma}_{\alpha\beta}^{\mathrm{I}}$ can be also obtained in a different manner. 
Writing Eq.~(\ref{eq:BMI}) in the eigenbasis, and simplifying the energy integration (see Appendix \ref{appa:bmI}) {the underlying kernel can be identified as 
\begin{equation}
\label{eq:bmIkernel}
\tilde{\mathcal{K}}^{\mathrm{I}}_{mn} = -\pi\hbar f'(\varepsilon_n) {\delta(\varepsilon_n - \varepsilon_m)},
\end{equation}
{which is an intraband like contribution. To recover the usual decay of the intraband response, one can approximate the Dirac delta in Eq.(\ref{eq:bmIkernel}) by a Lorentzian of the form $\delta(\varepsilon_n - \varepsilon_m)=\frac{1}{\pi}\frac{\eta}{\varepsilon_{nm}^2 + \eta^2}$. This results in a kernel which is  $\propto f'(\varepsilon_n)/\eta$ as found in ref. \cite{Huhtinen2023}. This same result can be obtained from the computation of the intraband limit of Eq. (\ref{eq:fullKernel}), as shown in  Appendix \ref{appa:intraband}.}

}

Equations (\ref{eq:generalResponse})-(\ref{eq:bmIkernel}) constitute the central result of this work. 

It is important to highlight that a relativistic correction to the St\v{r}eda decomposition was recently proposed \cite{Ado2024}. This introduces a third term, $\sigma_{\alpha\beta}^{\mathrm{III}}$, which can be expressed in a form analogous to $\sigma_{\alpha\beta}^{\mathrm{I}}$ and $\sigma_{\alpha\beta}^{\mathrm{II}}$:

\begin{equation}
    \sigma_{\alpha\beta}^{\mathrm{III}} = i\frac{e^2}{\hbar} \int d\varepsilon f(\varepsilon) \text{Tr} \left( [\boldsymbol{r}^{\alpha}, \boldsymbol{r}^{\beta}] \delta(\varepsilon - H) \right),
\end{equation}
where $\boldsymbol{r}^{\alpha}$ is the $\alpha$'s component of the position operator. Following our formulation one straightforwardly find:
\begin{equation}
    \sigma_{\alpha\beta}^{\mathrm{III}} = i\frac{e^2}{h} \sum_{n, \boldsymbol{k}} f(\varepsilon_n) [r^{\alpha}, r^{\beta}]_{nn}.
\end{equation}
While this term provides a correction to the Hall conductivity in the presence of relativistic spin-orbit coupling, its contribution will not be included in the numerical evaluations presented in this work.

\section{Numerical Implementation}

To validate our energy-integration-free expressions, we benchmark our analytical findings against established results in the literature. We consider a two-dimensional magnetic Rashba gas, which is often explored in similar contexts \cite{Bonbien2020, Ado2024}. The Hamiltonian for this system is given by:

\begin{equation}
\begin{split}
    H_{\mathrm{R}} = & -2\gamma (1 + \cos k_x + \cos k_y) + \Delta \sigma_z \\
    & + {\alpha} (\sin k_y \sigma_x - \sin k_x \sigma_y),
\end{split}
\end{equation}
where $\gamma$ is the hopping parameter, $\Delta$ represents the magnetic exchange energy, and ${\alpha}$ denotes the Rashba spin-orbit coupling strength. The operators $\sigma_x$ and $\sigma_y$ are the standard $2\times2$ Pauli matrices acting on the spin degree of freedom. By diagonalizing this Bloch Hamiltonian, we obtain the eigenvalues and eigenstates necessary to evaluate the transport responses.

Figure \ref{fig:fig1} illustrates the total Kubo response alongside its individual components. Our results show that the true sea term $\tilde{\sigma}_{xy}^{\mathrm{II}}$ accurately captures the Hall conductivity, while the surface response $\tilde{\sigma}_{xy}^{\mathrm{I}}$ vanishes, as expected for this regime. In contrast, the traditional Smr\v{c}ka-St\v{r}eda sea term $\sigma_{xy}^{\mathrm{II}}$ deviates from the total response due to the non-zero overlap term $\sigma_{\alpha\beta}^{ol}$. Nevertheless, the sum $\sigma_{xy}^{\mathrm{I}} + \sigma_{xy}^{\mathrm{II}}$ remains equal to the total response $\sigma_{xy}$.

These findings are in good agreement with the conclusions of Ref.~\cite{Bonbien2020}, where energy integrations were performed numerically. Furthermore, for the longitudinal conductivity, we find that the overlap term vanishes, therefore $\sigma_{\alpha\beta}^{\mathrm{II}}$ and $\tilde{\sigma}_{\alpha\beta}^{\mathrm{II}}$ become identical. 

{To further benchmark the validity of our formalism, a detailed analysis of the Fermi surface and Fermi sea contributions to the in-plane spin Hall current in a coplanar antiferromagnet is provided in the Supplemental Material. }
All numerical evaluations were performed using \texttt{Py4mulas} \cite{py4mulas}, a recently developed Python package for evaluating transport formulae in generic periodic systems. These two examples are available at \cite{py4mulasRashba, py4mulasKagome}.

\begin{figure*}
    \includegraphics[width=\textwidth]{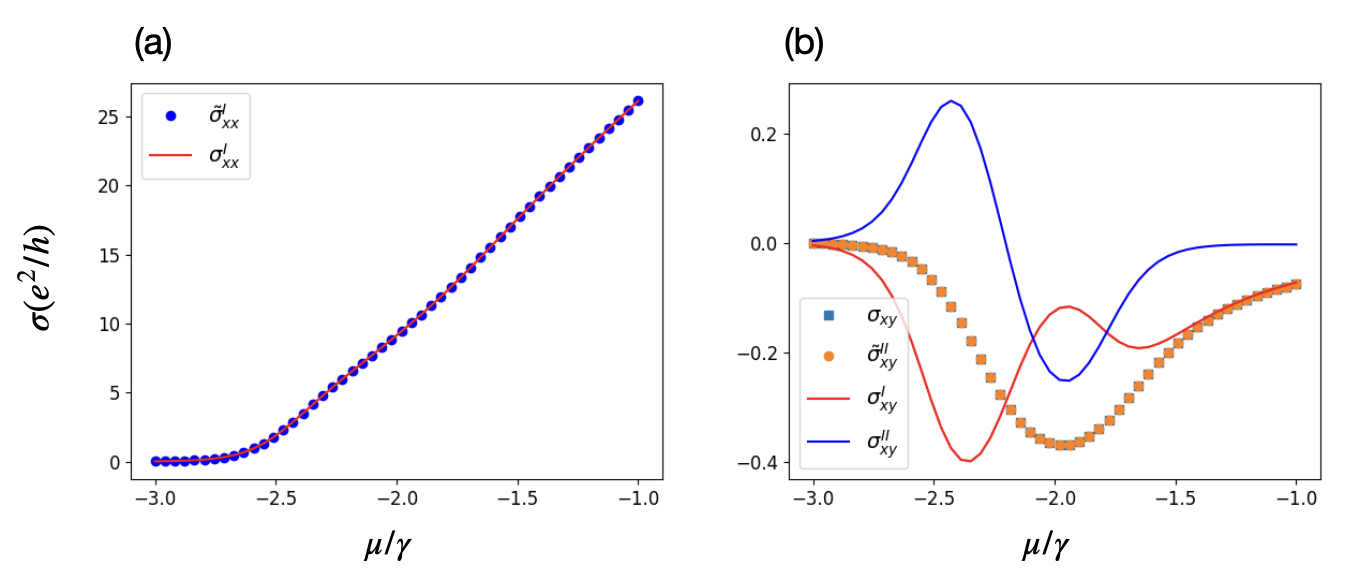}
    \caption{Transverse and longitudinal conductivity of the 2D magnetic Rashba gas. Panel (a) shows the longitudinal conductivity $\sigma_{xx}$ vs chemical potential $\mu$. Panel (b) shows different contributions of the transversal conductivity.
    The calculation is performed at $T=\eta=0.1\gamma$. The Hamiltonian parameters are taken as $\Delta=0.2\gamma$, ${\alpha}=1.4\gamma$. The conductivities are given in units of $\frac{e^2}{h}$. The momentum space summation is performed using $400\times400$ momentum vectors over the square Brillouin zone.}
    \label{fig:fig1}
\end{figure*}

\section{Discussion and Conclusions}
The numerical results presented in Fig. \ref{fig:fig1} demonstrate the robustness of our analytical reformulation. By comparing the standard Smr\v{c}ka-St\v{r}eda decomposition with the permutation-based surface and sea terms, we confirm that the overlap term $\sigma_{\alpha\beta}^{ol}$ is the primary source of the deviation between $\sigma_{\alpha\beta}^{\mathrm{II}}$ and the total Hall response $\sigma_{xy}$. Our eigenbasis-based approach, which bypasses numerical energy integration, captures these subtleties while significantly reducing the computational burden.

Specifically, for the magnetic Rashba gas, the fact that $\sigma_{xy} \approx \tilde{\sigma}_{xy}^{\mathrm{II}}$ highlights that the Hall response is dominated by the Fermi sea contribution, whereas the surface term vanishes in the absence of scattering. 
Our calculations of the longitudinal conductivity confirm that the overlap term is primarily relevant for Hall-like responses where the Berry curvature plays a central role \cite{Bonbien2020}. 

An advantage of the present formalism is that it drops the necessity of tuning energy integration bounds, which can sometimes be system dependent and/or requires particular adjustments for different transport scenarios.

Furthermore, it is important to emphasize that our formulation resolves the ambiguity associated with evaluating products of Green functions. In traditional approaches, the order in which the infinitesimal limits of the denominators are taken can lead to inconsistent results. This is particularly problematic in specific systems such as flat bands \cite{Huhtinen2023}, where standard St\v{r}eda terms have been found to yield unphysical results.

{
{Finally, we emphasize that while the present method offers a straightforward advantage for periodic systems by eliminating energy integration, its computational cost increases when considered for a real-space defined Hamiltonian.} Nevertheless, for intermediate-sized systems, bypassing the energy integration still provides a clear edge over alternative approaches like the Kernel Polynomial Method, which would still require extensive tuning of Chebyshev moments and precise handling of integration limits. However, for very large systems, the memory footprint required to diagonalize such real-space Hamiltonians can become prohibitive, rendering the method computationally inefficient at that scale.}

In summary, we have derived and implemented an energy-integration-free formulation of the Kubo-Bastin formula with its various decompositions. By transforming the traditional Green function expressions into the system's eigenbasis, we have demonstrated that the energy integrals can be performed analytically for generic systems in both real and momentum space. This approach offers a drastic increase in computational efficiency, particularly for large-scale systems (or large momentum space sampling) where dual momentum and energy integrations are traditionally prohibitive.

Our implementation, integrated into the \texttt{py4mulas} python package \cite{py4mulas}, provides a streamlined tool for cheaply evaluating linear transport responses in a wide range of condensed matter systems, with insightful physical separation between relevant transport components. 

{
\appendix
\section{Evaluation of The Energy Integrals}
\label{app:energy_integrals}

The energy integrations encountered in our calculations fall into two main categories depending on the number of Dirac-delta functions involved. 

The first type contains a single Dirac-delta function, appearing in equations such as Eqs.~(\ref{eq:kb}), (\ref{eq:sI}),  (\ref{eq:BMII}) and (\ref{eq:ol}). These integrals can be generalized in the form 
\begin{equation}
\label{eq:sigma1}
\Sigma_1 = \int d\varepsilon \, \Phi(\varepsilon) \, \text{Tr}\big[ \mathcal{O}^{\alpha} \mathcal{G}(\varepsilon) \mathcal{O}^{\beta} \delta (\varepsilon - H) \big],
\end{equation}
where the operator $\mathcal{G}(\varepsilon)$ represents either a Green function or its partial derivative with respect to energy, and $\Phi(\varepsilon)$ denotes either the Fermi-Dirac distribution function or its energy derivative, up to a multiplicative constant.

In the eigenbasis of the Hamiltonian, $\{|n\rangle\}$, with corresponding eigenvalues $\varepsilon_n$, the delta function expands as
\begin{equation}
\label{eq:delta_action}
\delta(\varepsilon - H) |n\rangle = \delta(\varepsilon - \varepsilon_n) |n\rangle.
\end{equation}
The Green function operator $\mathcal{G}(\varepsilon)$ shares the same eigenbasis representation and evaluates similarly as $\mathcal{G}(\varepsilon)|m\rangle = \mathcal{G}_m(\varepsilon)|m\rangle$.

By explicitly expressing the trace in Eq.~(\ref{eq:sigma1}) over this eigenbasis, we write
\begin{equation}
\label{eq:sigma1tr}
\Sigma_1 = \int d\varepsilon \, \Phi(\varepsilon) \sum_{n, \boldsymbol{k}} \langle n| \mathcal{O}^{\alpha} \mathcal{G}(\varepsilon) \mathcal{O}^{\alpha} \delta (\varepsilon - H) |n \rangle,
\end{equation}
where the sum over the momentum variable $\boldsymbol{k}$ is explicitly written. 
Inserting the resolution of identity $I = \sum_m |m\rangle\langle m|$ between $\mathcal{G}(\varepsilon)$ and $\mathcal{O}^{\alpha}$, and applying the spectral properties of both the delta function and the Green function, we obtain
\begin{equation}
\Sigma_1 = \int d\varepsilon \, \Phi(\varepsilon) \sum_{nm, \boldsymbol{k}} \mathcal{O}^{\alpha}_{nm} \mathcal{O}^{\alpha}_{mn} \mathcal{G}_m(\varepsilon) \delta (\varepsilon - \varepsilon_n),
\end{equation}
where $\mathcal{O}^{\alpha}_{nm} = \langle n|\mathcal{O}^{\alpha}|m\rangle$ and $\mathcal{O}^{\alpha}_{mn} = \langle m|\mathcal{O}^{\alpha}|n\rangle$. The Green function-like quantity $\mathcal{G}_m(\varepsilon)$ should be always a functional of $\varepsilon - \varepsilon_m$.  
Performing the integration over $\varepsilon$ we are left with the final expression:
\begin{equation}
\label{eq:sigma1final}
\Sigma_1 = \sum_{nm, \boldsymbol{k}} \Phi(\varepsilon_n) \mathcal{O}^{\alpha}_{nm} \mathcal{O}^{\beta}_{mn} \mathcal{G}_m(\varepsilon_n).
\end{equation}
Note that the quantity $\Phi(\varepsilon_n)\mathcal{G}_m(\varepsilon_n)$ corresponds to what we have defined as $\mathcal{K}_{mn}$ in Eq. (\ref{eq:generalResponse}).

The second type of integrals, encountered in Eq.~(\ref{eq:BMI}), contains a product of two Dirac-delta functions:
\begin{equation}
\label{eq:sigma2}
\Sigma_2 = \int d\varepsilon \, \Phi(\varepsilon) \, \text{Tr}\big[ \mathcal{O}^{\alpha} \delta (\varepsilon - H) \mathcal{O}^{\beta} \delta (\varepsilon - H) \big].
\end{equation}
Expanding the trace in the Hamiltonian eigenbasis and inserting a completeness relation between the operators results in
\begin{equation}
\label{eq:sigma2tr}
\Sigma_2 = \int d\varepsilon \, \Phi(\varepsilon) \sum_{nm, \boldsymbol{k}} \mathcal{O}^{\alpha}_{nm} \mathcal{O}^{\beta}_{mn} \delta (\varepsilon - \varepsilon_n) \delta (\varepsilon - \varepsilon_m).
\end{equation}
Integrating over $\varepsilon$ eliminates the first delta function by setting $\varepsilon = \varepsilon_n$ throughout the integrand, leaving behind the energy-conserving constraint on the remaining states:
\begin{equation}
\label{eq:sigma2final}
\Sigma_2 = \sum_{nm, \boldsymbol{k}} \Phi(\varepsilon_n) \mathcal{O}^{\alpha}_{nm} \mathcal{O}^{\beta}_{mn} \delta (\varepsilon_m - \varepsilon_n).
\end{equation}

{{{
\section{Evaluation of the Response Kernels}
\label{subsec:response_kernels}

To compute the explicit components of the conductivity tensor, we evaluate the response kernels $\mathcal{K}_{mn}$ associated with each term by adopting the general form introduced in Eqs.~(\ref{eq:sigma1final}) and (\ref{eq:sigma2final}). 

\subsection{Full Response Contribution}
\label{appa:full}
Comparing the full response described in Eq.~(\ref{eq:kb}) to the form in Eq. (\ref{eq:sigma1})), the operator $\mathcal{G}(\varepsilon)$ is identified as the energy derivative of the retarded Green function, $\mathcal{G}(\varepsilon) = \partial_{\varepsilon}G^r(\varepsilon) = -(G^r(\varepsilon))^2$. Taking the imaginary part of this expression yields directly the corresponding kernel (Eq. \ref{eq:fullKernel}):
\begin{equation}
\label{eq:full_kernel}
\mathcal{K}_{mn} = -2i\hbar \, f(\varepsilon_n) \frac{1}{(\varepsilon_{nm} + i\eta)^2},
\end{equation}
with $\varepsilon_{nm} = \varepsilon_n - \varepsilon_m$.

\subsection{Smr\v{c}ka-St\v{r}eda Contribution $\sigma^{I}$}
\label{appa:sI}
For the contribution $\sigma^{I}$ in Eq.~(\ref{eq:sI}), we take $\mathcal{G}(\varepsilon) = G^r(\varepsilon)$ and define the weight function as $\Phi(\varepsilon) = -i\hbar \, f'(\varepsilon)$ (according to Eq. (\ref{eq:sigma1})). This substitution directly yields the kernel formulated in Eq.~(\ref{eq:stredaIKernel}).

\begin{equation}
    \mathcal{K}^{I}_{mn} = -i\hbar f'(\varepsilon_{n})\frac{\varepsilon_{mn}}{\varepsilon_{mn} + i\eta}.
\end{equation}

\subsection{Fermi Sea Contribution}
\label{appa:bmII}
For the Fermi sea contribution in Eq.~(\ref{eq:BMII}), we set $\mathcal{G}(\varepsilon) = \partial_{\varepsilon}[G^r(\varepsilon) + G^a(\varepsilon)]$ with $\Phi(\varepsilon) = -i\hbar \, f(\varepsilon)$. A straightforward cyclic rearrangement of the operators within the trace ensures compatibility with the standard form of Eq.~(\ref{eq:sigma1}). 

Summing the derivatives of the retarded and advanced Green functions, the matrix elements evaluate to:
\begin{equation}
\label{eq:sea_g_elements}
\mathcal{G}_m(\varepsilon_n) = -\left[ \frac{1}{(\varepsilon_{mn} + i\eta)^2} + \frac{1}{(\varepsilon_{mn} - i\eta)^2} \right].
\end{equation}
This expression simplifies algebraically to:
\begin{equation}
\label{eq:sea_g_simplified}
\mathcal{G}_m(\varepsilon_n) = -2 \, \frac{\varepsilon_{mn}^2 - \eta^2}{(\varepsilon_{mn}^2 + \eta^2)^2}.
\end{equation}
Consequently, the Fermi sea kernel of Eq.~(\ref{eq:bmIIKernel}) is explicitly given by:
\begin{equation}
\label{eq:bmIIKernel_final}
\mathcal{K}^{II}_{mn} = \Phi(\varepsilon_n) \mathcal{G}_m(\varepsilon_n) = 2i\hbar \, f(\varepsilon_n) \frac{\varepsilon_{mn}^2 - \eta^2}{(\varepsilon_{mn}^2 + \eta^2)^2}.
\end{equation}

\subsection{Fermi Surface Contribution}
\label{appa:bmI}
The Fermi surface contribution Eq. (\ref{eq:BMI}) is written in the form $\Sigma_2$ given in Eq. (\ref{eq:sigma2tr}). Note that this form does not involve any Green function. But one can define an auxiliary $\mathcal{G}$ as $\mathcal{G}_m(\varepsilon_n)=\delta (\varepsilon_n - \varepsilon_m)$.  Choosing
$\Phi(\varepsilon)=-\pi\hbar f'(\varepsilon)$, 
the resulting kernel (Eq. (\ref{eq:bmIkernel})) is 

\begin{equation}
\tilde{\mathcal{K}}^{I}_{mn} = -\pi\hbar f'(\varepsilon_n)  \delta (\varepsilon_m - \varepsilon_n),
\end{equation}
which is an intraband like term.

\subsection{Overlap Contribution}
\label{appa:ol}
Comparing the overlap term Eq.(\ref{eq:ol}) with the form (\ref{eq:sigma1}), the weight function can be taken in the form of the energy derivative of the distribution function,
\begin{equation}
\label{eq:phiol}
    \Phi(\varepsilon) = -\frac{i\hbar}{2} f'(\varepsilon).
\end{equation} 
The $\mathcal{G}$ should be taken as 
$\mathcal{G}(\varepsilon) = G^r(\varepsilon) + G^a(\varepsilon)$. 

The corresponding matrix elements for $\mathcal{G}$ are therefore written as a direct sum of these Green functions:
\begin{equation}
\label{eq:overlap_g}
\mathcal{G}_m(\varepsilon_n) = \frac{1}{\varepsilon_{mn} + i\eta} + \frac{1}{\varepsilon_{mn} - i\eta},
\end{equation}
which reduces to:
\begin{equation}
\label{eq:overlap_g_simplified}
\mathcal{G}_m(\varepsilon_n) = \frac{2\varepsilon_{mn}}{\varepsilon_{mn}^2 + \eta^2}.
\end{equation}
Multiplying by $\Phi(\varepsilon_n)$ (Eq. (\ref{eq:phiol})) yields the final overlap kernel presented in Eq.~(\ref{eq:olKernel}),
\begin{equation}
    \mathcal{K}^{ol}_{mn} = -i\hbar f'(\varepsilon_{n})\frac{\varepsilon_{mn}}{\varepsilon_{mn}^2 + \eta^2}.
\end{equation} 

Finally, we remark that the energy integration in the second Smr\v{c}ka-St\v{r}eda contribution $\sigma ^{II}$ (Eq. (\ref{eq:sII})) remains unreduced. The absence of an explicit Dirac delta function prevents the application of our integration-free technique to bypass the energy variable. However, this component can still be determined with minimal computational cost by employing the remaining contributions as outlined in Eq.~(\ref{eq:fromOverlap}).

}}}

\section{Intraband contribution of the Kubo-Bastin formula}
\label{appa:intraband}
In Eq. (\ref{eq:fullKernel}), we have shown the full kernel of the Kubo-Bastin formula. Yet it is important to show how the intraband response behaves at a given broadening $\eta$.
To do so, we start from the general form of the responses as given in (\ref{eq:generalResponse})

\begin{equation}
    \label{eq:generalResponseAppa}
    \sigma_{\alpha\beta}=\sum_{mn, \boldsymbol{k}} \mathcal{O}^{\beta}_{mn}(\boldsymbol{k}) \mathcal{O}^{\alpha}_{nm}(\boldsymbol{k})  \mathcal{K}_{mn}(\boldsymbol{k}).
\end{equation}
Note that we are interested in the real part of this quantity.

So, let us write it as 
\begin{equation}
    \rm{Re} \sigma_{\alpha\beta} = \frac{1}{2} (\sigma_{\alpha\beta} + \sigma_{\alpha\beta}^*).
\end{equation}
This leads naturally to 
\begin{align}
    2\rm{Re} \sigma_{\alpha\beta} &=  \sum_{mn, \boldsymbol{k}} \mathcal{O}^{\beta}_{mn}(\boldsymbol{k}) \mathcal{O}^{\alpha}_{nm}(\boldsymbol{k})  \mathcal{K}_{mn}(\boldsymbol{k}) \\
    &\quad + \sum_{mn, \boldsymbol{k}}  \mathcal{O}^{\beta *}_{mn}(\boldsymbol{k}) \mathcal{O}^{\alpha *}_{nm}(\boldsymbol{k}) \mathcal{K}^*_{mn}(\boldsymbol{k}).
\end{align}
Assuming  $\mathcal{O}^{\beta}$ and $\mathcal{O}^{\alpha}$ are hermitian operators, we have $\mathcal{O}^{\beta *}_{mn}=\mathcal{O}^{\beta}_{nm}$.
Interchanging the indices in the second sum one can write
\begin{equation}
    2\rm{Re} \sigma_{\alpha\beta} =  \sum_{mn, \boldsymbol{k}} \mathcal{O}^{\beta}_{mn}(\boldsymbol{k}) \mathcal{O}^{\alpha}_{nm}(\boldsymbol{k})  (\mathcal{K}_{mn}(\boldsymbol{k}) +  \mathcal{K}^*_{nm}(\boldsymbol{k})).
\end{equation}
We technically prefer to write the kernel in Eq.(\ref{eq:full_kernel}) as
\begin{equation}
\label{eq:full_kernel2}
\mathcal{K}_{mn} = -2i\hbar \, f(\varepsilon_n) \frac{1}{\varepsilon_{nm}(\varepsilon_{nm} + i\eta)}.
\end{equation}
In fact, in  (\ref{eq:full_kernel}) we took $\partial_{\varepsilon} G^r(\varepsilon) = -(G^r(\varepsilon))^2$.
However, it is sometimes technically more convenient  to 
rather take (see Eq. (A7) of ref. \cite{Crepieux2001})
\begin{equation}
    \lim_{\eta\rightarrow 0}\partial_{\varepsilon} G^r(\varepsilon) = \lim_{\eta\rightarrow 0}\frac{1}{\varepsilon_{nm} (\varepsilon_{nm} + i\eta)},
\end{equation}
though both limits are equivalent.
}
From this we find 
\begin{equation}
    \mathcal{K}_{mn}(\boldsymbol{k}) +  \mathcal{K}^*_{nm}(\boldsymbol{k}) = -2i\hbar\frac{f(\varepsilon_{n}) - f(\varepsilon_{m})}{\varepsilon_{nm} (\varepsilon_{nm} +i\eta)}.
\end{equation}
Now, taking the limit $\varepsilon_{nm}\rightarrow 0$, we obtain
\begin{equation}
    \mathcal{K}_{mn}(\boldsymbol{k}) +  \mathcal{K}^*_{nm}(\boldsymbol{k}) = -2i\hbar \lim_{\varepsilon_{nm}\rightarrow 0}\frac{f(\varepsilon_{n}) - f(\varepsilon_{m})}{\varepsilon_{n} - \varepsilon_{m} }\frac{1}{i\eta}.
\end{equation}
Further, this sum reduces to 
\begin{equation}
    \mathcal{K}_{mn}(\boldsymbol{k}) +  \mathcal{K}^*_{nm}(\boldsymbol{k}) = - \frac{2\hbar}{\eta} \partial_{\varepsilon} f(\varepsilon)|_{\varepsilon=\varepsilon_n}.
\end{equation}
Finally, the response $\sigma_{\alpha\beta}$ becomes

\begin{equation}
    \sigma_{\alpha\beta} =  - \frac{\hbar}{\eta}\sum_{n, \boldsymbol{k}} \mathcal{O}^{\beta}_{nn}(\boldsymbol{k}) \mathcal{O}^{\alpha}_{nn}(\boldsymbol{k}) \partial_{\varepsilon} f(\varepsilon)|_{\varepsilon=\varepsilon_n},
\end{equation}
which is the well known Drude like response, inversely proportional to the broadening parameter $\eta$ \cite{Huhtinen2023}.

One can actually similarly show that the Fermi-sea  term (Eq. (\ref{eq:BMII})) is purely intrband and has no intraband contribution. This cancellation can be directly spotted from the structure of the underlying $\mathcal{G}$ function in Eq. (\ref{eq:sea_g_elements}), where the first term of the right hand side of such equation comes with a contribution $\propto 1/i\eta$, while the second comes with a contribution $\propto -1/i\eta$. This means that in the decomposition terms only the Fermi-surface component (Eq. (\ref{eq:BMI})) has an intraband contribution, as discussed below Eq. (\ref{eq:bmIkernel}).

\bibliography{refs}
\end{document}